\documentclass[11pt]{article}
\usepackage{amsmath,amssymb,bm}
\usepackage{graphics}
\usepackage{graphicx}
\usepackage{amssymb}
\usepackage{epstopdf}

\def\herwig{{\sc Herwig}}

\usepackage{mystyle}
\usepackage{cite,./mcite}
\usepackage{authblk}
\usepackage{lineno}
                \def\lsim{\mathrel{\rlap{\lower4pt\hbox{\hskip1pt$\sim$}}
    \raise1pt\hbox{$<$}}}                \def\gsim{\mathrel{\rlap{\lower4pt\hbox{\hskip1pt$\sim$}}
    \raise1pt\hbox{$>$}}}

\def\MC{MCEG}

\title{Summary of the Workshop on Multi-Parton Interactions (MPI@LHC 2012) }
\author[1]{H. Abramowicz\footnote{partly supported by the Israel Science Foundation}}
\author[2]{P. Bartalini}
\author[3]{M. B\"ahr}
\author[4]{N. Cartiglia}
\author[5]{R. Ciesielski}
\author[6]{E. Dobson}
\author[7]{F. Ferro}
\author[5]{K. Goulianos}
\author[8]{B. Guiot}
\author[9]{X. Janssen} 
\author[10]{H. Jung}
\author[11]{Iu. Karpenko}
\author[12]{J. Kaspar} 
\author[10]{J. Katzy} 
\author[13]{F. Krauss} 
\author[14]{P. Laycock} 
\author[15]{E. Levin}
\author[16]{M. Mangano} 
\author[5]{Ch. Mesropian}  
\author[17]{A. Moraes}
\author[18]{M. Myska}
\author[19]{D. Moran}
\author[20]{R. Muresan}
\author[10]{Z. Nagy}
\author[21]{T. Pierog}
\author[6]{A. Pilkington} 
\author[22]{M. Poghosyan} 
\author[23]{T. Rogers}
\author[24]{S. Sen}
\author[25]{M.H. Seymour}
\author[25]{A. Siodmok}
\author[26]{M. Strikman}
\author[15]{P. Skands}
\author[27]{D. Treleani}
\author[28]{D. Volyanskyy} 
\author[8]{K. Werner}
\author[6]{P. Wijeratne}
{\tiny
\affil[1]{Raymond and Beverly Sackler Faculty of Exact Sciences, School of Physics, Tel Aviv University, Israel }
\affil[2]{National Taiwan University (NTU), Taipei, Taiwan}
\affil[3]{Blue Yonder GmbH \& Co. KG, Karlsruher Strasse 88, 76139 Karlsruhe, Germany}
\affil[4]{INFN Sezione di Torino a, Universita di Torino b, Universita del Piemonte Orientale (Novara) c, Torino, Italy}
\affil[5]{The Rockefeller University, New York, USA}
\affil[6]{Department of Physics and Astronomy, University College London,  United Kingdom }
\affil[7]{Sezione INFN di Genova - Genova, Italy, EU }
\affil[8]{SUBATECH, University of Nantes, IN2P3/CNRS, EMN, Nantes, France}
\affil[9]{Universiteit Antwerpen,  Belgium}
\affil[10]{Deutsches Elektronen-Synchrotron DESY, Hamburg, Germany }
\affil[11]{Bogolyubov Institute for Theoretical Physics, Kiev, Ukraine }
\affil[12]{Institute of Physics of the Academy of Sciences of the Czech Republic, Praha, Czech Republic }
\affil[13]{Institute for Particle Physics Phenomenology, Durham University, UK }
\affil[14]{Oliver Lodge Laboratory, University of Liverpool, United Kingdom }
\affil[15]{ Departamento de Fisica, Universidad Tecnica Federico Santa Maria and Centro Cientifico-Tecnologico de Valparaiso, Chile and 
Department of Particle Physics, School of Physics and Astronomy, Tel Aviv University, Israel}
\affil[16]{Department of Physics, CERN, Theory Unit, Geneva, Switzerland }
\affil[17]{SUPA - School of Physics and Astronomy, University of Glasgow, United Kingdom }
\affil[18]{FNSPE, Czech Technical University in Prague,  Czech Republic }
\affil[19]{School of Physics and Astronomy, University of Manchester, United Kingdom}
\affil[20]{Horia Hulubei National Institute of Physics and Nuclear Engineering, Bucharest-Magurele, Romania and Petersburg Nuclear Physics Institute (PNPI), Gatchina, Russia }
\affil[21]{Karlsruhe Inst. of Technology, KIT, Campus North, Inst. f. Kernphysik, Germany}
\affil[22]{European Organization for Nuclear Research (CERN), Geneva, Switzerland}
\affil[23]{C.N. Yang Institute for Theoretical Physics, Stony Brook University, USA}
\affil[24]{The University of Iowa, USA }
\affil[25]{Consortium for Fundamental Physics, School of Physics and Astronomy, The University of Manchester, United Kingdom}
\affil[26]{The Pennsylvania State University, University Park, PA 16802, USA}
\affil[27]{Dipartimento di Fisica dell Università di Trieste and INFN, Sezione di Trieste, Trieste, Italy}
\affil[28]{Max-Planck-Institut f\"ur Kernphysik (MPIK), Heidelberg, Germany}
} 
\begin{document}

\maketitle 

\section{Introduction}
The increasing interest in Multiple Parton Interactions (MPI) studies resulted in a first workshop on MPI at the LHC held in Perugia in October 2008. The unique feature of this workshop, which had a second edition in Glasgow (December 2010), a third at DESY (November 2011) and a fourth at CERN (December 2012) was to bring together different communities interested in MPI coming from different areas. 
Typically Monte Carlo event generators (\MC s) are used in the description of multiparton interactions, however, in the last years several theoretical investigations were published which were also discussed during this workshop.

The workshop MPI@LHC 2012 was structured in four working groups:
\begin{enumerate}
\item MPI and diffraction \\
{\it Working Group Conveners:} \\ M. Poghosyan, A. Pilkington, N. Cartiglia, D. Volyanskyy, F. Ferro
\item MPI and small-x \\
{\it Working Group Conveners: } \\ H. Abramowicz, H. Jung 
\item MPI and double-parton scattering\\
{\it Working Group Conveners: } \\
 E. Dobson, A. Moraes, P. Bartalini, X. Janssen, R. Muresan, D. Treleani
\item MPI and MC development \\
{\it Working Group Conveners: } \\ J. Katzy, F. Krauss, Z. Nagy, T. Rogers \\
\end{enumerate}

The slides presented at MPI@LHC 2012 can be found in \cite{mpilhc2012}.
In the following sections we document with short resumes and highlights the discussions in the different working groups.
In the appendix extra contributions to this summary are included.

\section{MPI and diffraction}
\makeatletter{}{\it Authors: }\\
N.~Cartiglia, F.~Ferro, J.~Kaspar, P.~Laycock, E~ Levin,  
Ch.~Mesropian, D.~Moran, A.~Pilkington, M.~Poghosyan,  
S.~Sen, P.~Wijeratne, D.~Volyanskyy 
\\ 

This section presents a brief review of the experimental results presented during the {\it MPI and  Diffraction} Working Group session. Outstanding questions and opportunities for future measurements are discussed.

\subsection{Summary of experimental results}
LHCb presented measurements of forward energy flow in $pp$ collisions at $\sqrt{s} = 7 $ TeV for inclusive 
interactions, hard scattering processes and events with enhanced/suppressed diffractive contribution. None of the \MC\  models were able to describe the data for all event classes studied. Measurements of exclusive $J/\psi$ and $\psi(2S)$ vector meson production were also presented in the dimuon decay channel, showing good agreement with previous experimental measurements and theoretical predictions. The dependency of the $J/\psi$ photo-production cross-section on the $\gamma p$ centre-of-mass energy was presented, showing consistency with previous HERA measurements. 

ATLAS presented measurements of the transverse energy flow as a function of pseudorapidity $|\eta|$ in $pp$ collisions at centre-of-mass energy of $\sqrt{s} = 7 $ TeV for quasi-inclusive $pp$ collisions and dijet topologies. All the default \MC\  tunes produced too little activity at the largest pseudorapidities. The choice of parton density function was observed to have a large impact on the \MC\  predictions. The dependence of the inelastic cross section  on the size of a forward pseudorapidity gap (defined as no particles with  $p_{T}> 200$~MeV in a given $\eta$-interval) was also presented. None of the \MC\  models reproduced the rise of the cross section at large gap-sizes observed in the data. The \herwig ++ \MC\  produced a diffractive-like spectrum despite not containing a diffractive component in its minimum bias model.

CMS presented the differential cross section for dijet production in the pseudorapidity region $|\eta^{j1,j2}| < 4.4$ with $p_{T}^{j1,j2} > 20$ GeV as a function of $\xi$, which approximates to the fractional momentum loss of the scattered proton in single-diffraction events. Single diffractive (SD) events are observed to dominate at low $\xi$. The rapidity gap survival probability for SD events is extracted from the data to be $0.08 \pm 0.04$ (NLO) or $0.12 \pm 0.05$ (LO). Furthermore, the presence of a diffractive component was demonstrated in events containing $W/Z$ bosons in addition to a large pseudorapidity gap;  the diffractive component for the CMS analysis cuts was found to be $50.0 \pm 9.3(\rm stat.) \pm 5.2(\rm syst.)$~\%.

The TOTEM experiments has measured the elastic, inelastic and total cross-sections of proton-proton collisions at the centre-of-mass energy of $\sqrt s = 7\,\rm TeV$\cite{Antchev:2013gaa,Antchev:2013iaa}. All these cross-sections have been determined in three
complementary ways, exploiting data from the very forward proton detectors and/or the more central charged-particle telescopes. A luminosity-independent measurement has been repeated at the energy of $\sqrt{s} = 8$~TeV \cite{Antchev:1495764}. These measurements exploit data from the very forward proton detectors and/or the more central charged-particle telescopes. Both $7$ and $8\,\rm TeV$ total cross-section measurements (with uncertainties of about $2.5\,\%$) are consistent with extrapolations of lower energy data from the COMPETE collaboration\cite{Cudell:2002xe}.

The H1 and ZEUS collaborations presented an update on their diffractive measurements, with diffractive events selected either by tagging the scattered proton or by identifying events with a large rapidity gap (LRG) in the central detector.  Both collaborations have analyzed their diffractive DIS data using NLO QCD, after showing that proton vertex factorization holds within the experimental uncertainties.  The QCD fits are able to simultaneously describe and/or fit the inclusive diffractive DIS data and dijet data, as well as accurately predicting $F_L^D$. The H1 and ZEUS LRG data show good agreement over a large kinematic range.  H1 has also published an analysis of diffractive DIS data taken with lower proton beam energies, allowing the simultaneous extraction of both $F_2^D$ and the longitudinal diffractive structure function, $F_L^D$. 
The final proton-tagged data from both experiments have been combined to provide the first HERA diffractive dataset.

\subsection{Future outlook and outstanding questions}

The experimental results presented at this workshop were an impressive start to a diffractive program at the LHC. The experiments also presented their plans for future measurements. Nevertheless, issues and questions were highlighted during the discussion and the working group encourage the experiments and theorists to follow up on the following key issues:
\begin{itemize}
\item The \herwig ++ hadronization model produces a diffractive-like rapidity gap spectra from an MPI model. Can this model be expanded in any way to produce a new type of diffractive model? Hadronization models may turn out to limit the scope of future measurements. The new {\sc Sherpa} model may be helpful in this respect.
\item It would be nice to understand the connection between MPIs and diffraction, both at the fundamental level, and at the level of \MC\  modeling.
\item Both ATLAS and TOTEM/CMS can perform the standard soft-QCD measurements, using the central tracking and calorimetry, with the added requirement of two forward protons to select Double-Pomeron-Exchange style events. This would allow the entire soft-QCD program to be repeated to determine the similarities and differences between inclusive and diffractive hadron production mechanisms.
\item Precision measurements of diffractive cross sections are needed for multiple processes. This will allow the theoretical models to be scrutinized and the rapidity gap suppression (soft-survival) factor to be extracted from the data. Efforts to reduce the experimental and theoretical systematic uncertainties associated with soft-survival factor extraction would be useful to allow a detailed comparison of soft-survival factor between different diffractive processes.
\item The LHCb and ATLAS forward energy flow data can be used to constrain the UE modeling in the forward region. Performing forward energy flow measurements at ATLAS/CMS/LHCb in a common fiducial region would also be of great help in interpreting the data.
\item Data from ALFA is crucial to confirm the current total cross section measurements by TOTEM. The current TOTEM results, when taken with the fiducial inelastic cross section measurements by ATLAS/CMS/ALICE, indicate that the contribution from low-mass diffraction is larger than many (but not all) of the theoretical or Monte Carlo models. The combined data from LHC experiments can be used to place strong constraints on the models of low-mass diffraction.
\item In the same spirit, the contribution from low-mass single diffraction can be measured directly using forward proton tagged events at ALFA or TOTEM/CMS.
\end{itemize}

\section{MPI and small-x}
{\it Authors:} \\
H. Abramowicz, H. Jung\\

In the {\it MPI and small $x$} session, L.~L\"onnblad discussed the similarity of processes simulated as multi-parton interactions including DGLAP parton evolution processes compared to processes simulated according to BFKL/CCFM evolution. F.~Hautmann discussed processes simulated using transverse momentum dependent parton distributions convoluted with appropriate matrix elements giving predictions similar to what is obtained with MPI simulation.

In the experimental part G. Grindhammer (H1) showed measurements from $ep$ scattering which are sensitive to multiparton interactions.
C. Bourdarios (ATLAS) showed measurements of transverse energy flow extending to the forward region and stressed that the measurements of event shape variables are complementary  to the usual charged particle based measurements to investigate the underlying event.
I. Hos (CMS) presented the different forward jet cross section measurements and showed how large MPI processes contribute.  P. Spradlin (LHCb) showed new measurements of forward production of Drell-Yan $c\bar{c}$ and $b\bar{b}$ production and a comparison with theoretical predictions including MPI.

In the following we briefly illustrate the connection of multiparton interaction and small $x$ physics.
In the collinear factorization ansatz a cross section $A+B \to X$ is described as a convolution of the parton density $f(x,\mu^2)$ with a partonic cross section $\hat{\sigma}(i+j\to X)$:
\begin{equation}
\sigma= f^A_i(x_1,\mu^2) \hat{\sigma}(i+j\to X) f^B_j(x_2,\mu^2) +{\cal O}(1/\mu^2)
\end{equation}
This factorization ansatz is proven for a limited number of processes and is valid up to corrections  which are power suppressed by ${\cal O}(1/\mu^2)$.

In general a cross section at highest energies, as obtained at the LHC, can be written as \cite{Manohar:2012pe}  (leaving out for simplicity flavor mixing terms):
\begin{equation}
\sigma \sim   c_1 f_q f_q + c_2 f_g f_g + c_3 f_g F + c_4 F F  
\end{equation}
where $c_1, c_2, c_3, c_4$ are the partonic cross sections and $f(F)$ are the parton densities for single (multiple) parton interactions. The first two terms describe the interaction via one parton exchange (quark and/or gluon, as in eq.(1)), while the last two terms allow for more than one parton exchange (double parton exchange), which are suppressed by powers of $\mu$  compared to the leading terms $c_1, c_2$. These power suppressed terms are by definition not included in collinear factorization, and have to be added, leading to the introduction of double parton scattering (DPS) or MPI.

Two different scenarios appear, where DPS or more generally MPI can be important: at small $x$, where the parton densities are very large, the terms $c_3, c_4$ receive extra contributions from $F\sim f^2$, and can become of similar size as the terms $c_1, c_2$. If, on the other hand, by some dedicated selection rules, the terms $c_1, c_2$ are suppressed, then again the contribution from DPS can become of similar size as the single parton exchange process. 

At small $x$, where the parton densities are large, collinear factorization is expected not to be sufficient: the inclusion of BFKL (or CCFM) effects into the parton evolution seems to be necessary. At small $x$, the splitting functions in the evolution equations receive contribution of the type $k_t/q_t$: 
\begin{equation}
{\cal P}_{g q}(z,q_t,k_t) = P_{qg,DGLAP}(z)\left ( 1 + \sum_{n=0}^{\infty} b_n(z)(k_t^2/q_t^2)^n \right)
\end{equation}
With this improved splitting function, terms which are neglected in collinear factorization are included. 

It has been shown, that the application of CCFM leads to final state predictions which go beyond what is obtained from simulations using DGLAP parton showers and is in some regions similar to what is obtained including MPI effects.
The similarity of final states generated by more sophisticated parton-splitting mechanisms and by MPIs warrants further investigation.

\section{MPI and double-parton scattering}
 
\makeatletter{}

\def\n {\nonumber}
\def\a {\epsilon}
\def\g {\gamma}
\def\M  {{\cal M}}
\def\F  {{\cal F}}
\def\O  {{\cal O}}
\def\R  {{\cal R}}
\def\N  {{\cal N}}
\def\J {$J/\psi$ }
\def\Js {$J/\psi$'s }
\def\j { J/\psi  }
\def\pjt {$p_{\psi,T}$ }
\def\cc {$c\bar{c}$ }
\def\cj {$\chi_{cJ}$ }
\def\bb {$b\bar{b}$ }
\def\qq {$Q\bar{Q}$ }
\def\qe {quasielastic}
\def\pd {pseudo-diffractive}
\def\ktf {$k_t$-factorization }
\def\ktfa {$k_t$-factorization approach }
\newcommand{\bpsi}{\mbox{\boldmath $\psi$}}
\newcommand{\bPhi}{\mbox{\boldmath $\Phi$}}
\newcommand{\bkappa}{\mbox{\boldmath $\kappa$}}
\newcommand{\bl}{\mbox{\boldmath $l$}}
\newcommand{\bp}{\mbox{\boldmath $p$}}
\newcommand{\bq}{\mbox{\boldmath $q$}}
\newcommand{\br}{\mbox{\boldmath $r$}}
\newcommand{\bs}{\mbox{\boldmath $s$}}   
\newcommand{\bk}{\mbox{\boldmath $k$}}
\newcommand{\bfb}{\mbox{\boldmath $b$}}
\newcommand{\bE}{\mbox{\boldmath $E$}}
\newcommand{\Jot}{{\cal{J}}}
\def\cpc#1#2#3  {{Computer\ Phys.\ Comm.\ }  {\bf#1}, #2 (#3)}
\def\err#1#2#3  {{\it Erratum }              {\bf#1}, #2 (#3)}
\def\epjc#1#2#3 {{Eur. Phys. J. C }          {\bf#1}, #2 (#3)}
\def\dum#1#2#3  {{~}                         {\bf#1}, #2 (#3)}
\def\ib#1#2#3   {{\it ibid. }                {\bf#1}, #2 (#3)}
\def\jcp#1#2#3  {{J.\ Comput.\ Phys.\ }      {\bf#1}, #2 (#3)}
\def\jetpl#1#2#3 {{\rm JETP Lett.}           {\bf#1}, #2 (#3)}
\def\jhep#1#2#3 {{JHEP }                     {\bf#1}, #2 (#3)}
\def\ijmp#1#2#3 {{Int.\ J.\ Mod.\ Phys.\ }   {\bf#1}, #2 (#3)}
\def\jpg#1#2#3  {{J.\ Phys.\ G }             {\bf#1}, #2 (#3)}
\def\mpl#1#2#3  {{Mod.\ Phys.\ Lett.\ }      {\bf#1}, #2 (#3)}
\def\ncim#1#2#3 {{Nuovo Cimento }            {\bf#1}, #2 (#3)}
\def\np#1#2#3   {{Nucl.\ Phys.\ }            {\bf#1}, #2 (#3)}
\def\npb#1#2#3  {{Nucl.\ Phys.\ B}           {\bf#1}, #2 (#3)}
\def\pan#1#2#3  {{Phys.\ At.\ Nuclei }       {\bf#1}, #2 (#3)}
\def\plb#1#2#3  {{Phys.\ Lett.\ B }          {\bf#1}, #2 (#3)}
\def\prep#1#2#3 {{Phys.\ Rep.\ }             {\bf#1}, #2 (#3)}
\def\prd#1#2#3  {{Phys.\ Rev.\ D }           {\bf#1}, #2 (#3)}
\def\prl#1#2#3  {{Phys.\ Rev.\ Lett.\ }      {\bf#1}, #2 (#3)}
\def\ptp#1#2#3  {{Prog.\ Theor.\ Phys.\ }    {\bf#1}, #2 (#3)}
\def\ps#1#2#3   {{Physica Scripta }          {\bf#1}, #2 (#3)}
\def\rmp#1#2#3  {{Rev.\ Mod.\ Phys.\ }       {\bf#1}, #2 (#3)}
\def\rpp#1#2#3  {{Rep.\ Prog.\ Phys.\ }      {\bf#1}, #2 (#3)}
\def\sa#1#2#3   {{Sci. Acta}                 {\bf#1}, #2 (#3)}
\def\sjnp#1#2#3 {{Sov.\ J.\ Nucl.\ Phys.\ }  {\bf#1}, #2 (#3)}
\def\spj#1#2#3  {{Sov.\ Phys.\ JETP }        {\bf#1}, #2 (#3)}
\def\spjl#1#2#3 {{Sov.\ JETP Lett.\ }        {\bf#1}, #2 (#3)}
\def\spu#1#2#3  {{Sov.\ Phys.-Usp.\ }        {\bf#1}, #2 (#3)}
\def\yaf#1#2#3  {{Yad.\ Fiz.\ }              {\bf#1}, #2 (#3)}
\def\zp#1#2#3   {{Zeit.\ Phys.\ }            {\bf#1}, #2 (#3)}
\def\zpc#1#2#3  {{Z.\ Phys.\ C }             {\bf#1}, #2 (#3)}
\def\etal {{\it et al. }}

\subsection{Experimental Discussion}
{\it Authors: }\\
P. Bartalini, R. Mure\c san\\

This section is a summary focusing on the experimental contributions in the context of the Double Parton Scattering (DPS) session.

Experts from the QCD, heavy flavors, Electro-Weak, top, diffraction, nuclear and cosmic rays fields, with almost an equal number of experimentalists and theorists were participating in this session. Phenomenologists, authors of the general purpose Monte Carlo generators for particle and nuclear physics, the latter covering also the field of cosmic rays were contributing. 

\subsubsection{State of the art}

The status of the DPS theory is briefly discussed in the theoretical summary of the DPS session; here we assume the same formalism. The expression of the inclusive DPS cross section, $\sigma_{double}^{(A,B)}$, used is:
\begin{eqnarray}
\sigma_{double}^{(A,B)}=\frac{m}{2}\frac{\sigma_A\sigma_B}{\sigma_{eff}}
\label{sdouble}
\end{eqnarray}
where $A$ and $B$ represent the two elementary partonic processes of interest, taking place in two disconnected regions of phase space, $m$ is a symmetry factor whose value is $m=1$ if $A$ and $B$ are identical processes and $m=2$ if $A$ and $B$  are different processes. $\sigma_A$ and $\sigma_B$ are the two inclusive cross sections for observing either the process $A$ or the process $B$  in the same hadronic collision, while all unknowns converge in the value of a single quantity with the dimensions of a cross section, $\sigma_{eff}$. 

The typical signature of a DPS event is the lack of angular and/or transverse momentum, $p_T$, correlations between the kinematic properties of the two individual scatterings.

The production of four-jets of high  $p_T$  is the most prominent signature for the multiple high  $p_T$ scatterings at hadron colliders as it may involve two independent scatters in the same collision, each of them producing a jet pair. As opposed to the naive assumptions based on uncorrelated scatterings, which would basically imply $\sigma_{eff} = \sigma_{inel}$, the early DPS measurements~\cite{Akesson:1986iv,Abe:1993rv} that inspired the first MPI models~\cite{Sjostrand:1986ep} seem to favor much smaller values of $\sigma_{eff}$, which would indicate a DPS enhancement.

However, the measurements of $\sigma_{eff}$ in four-jet events have to face the major difficulty of vetoing the significant background coming from other sources of jet production, in particular from the QCD bremsstrahlung. On top of that, a simplified description of  the combinatorial background was used focusing on DPS topologies with individual  interactions characterized by different objects in the final state or different kinematics. 
This is why the Tevatron strategy to measure directly the hard MPI rates relies mostly on the reconstruction of extra jet pairs in final states with direct photons~\cite{Abe:1997xk,Abazov:2009gc}.
This method also exploits the better energy resolution performances of photons with respect to jets, resulting in a significant reduction of the combinatoric background from the two interactions. The  $\sigma_{eff}$ measurements in the \mbox{3 jet + $\gamma$} channel confirm the early observations in the 4-jets channel.

The CDF measurement of double parton scattering~\cite{Abe:1997xk} uses a non-standard 
definition of $\sigma_{eff}$ that makes this important quantity process dependent. 
This has been pointed out and corrected for the first time in Ref.~\cite{Treleani:2007gi}.  In section \ref{bahr} the CDF's event definition is used in order to provide an improved correction.

The re-interpretation of the Tevatron DPS phenomenology should be encouraged not only regarding the extraction of standard observables which can be directly compared with the LHC measurements, indeed the conceptual progress made by the theory should also be taken into account. The single parton scattering (SPS) background modeling in the Tevatron DPS analyses does often follow a trivial approach, which is no longer justified in the light of the modern MC tools, which can account for the higher order corrections. The importance of refining the SPS description in the Tevatron measurements emerged  from the  MPI workshop discussions. 

\subsubsection{LHC studies}

Given the uncertain DPS legacy from previous colliders it is even more 
important to set up a robust research program on the study of hard MPI at 
the LHC experiments.

The first LHC results with big impact in the light of DPS  phenomenology were 
provided by  LHCb with the papers reporting on double 
$J/\psi$~\cite{Aaij:2011yc},  double open charm and open charm $J/\psi$ 
production~\cite{Aaij:2012dz}. These two papers together provide for the  
first time eleven charm  pair production processes measured in hadron collisions, 
exploiting the unique fiducial coverage of the LHCb experiment. The 
$J/\psi$-pair production measurement result is in good agreement with 
LO color singlet QCD prediction~\cite{Kiselev:1988mc,Qiao:2009kg}, being at the same time considered by theorists as consistent with about ~30\% contribution coming from the DPS processes \cite{Novoselov:2011ff}. In contrast with this first result showing a good agreement both with the SPS/pQCD and DPS interpretations, the measured cross-section of the $J/\psi$ produced in association with a open charm hadron, as well as the double open charm production, strongly disagree with SPS/pQCD interpretation and support a significant DPS contribution, appearing to be a factor of 20-60 larger than LO QCD calculations \cite{Berezhnoy:1998aa}, see Fig. 9 from \cite{Aaij:2012dz} and the quoted theory prediction. 
%DPS is only partially supported by double open charm production results, since the measured effective cross section is a  factor of 2-3 larger than what is obtained for double open charm-processes. 
DPS is only partially supported by double open charm production results, see Fig. 10 in \cite{Aaij:2012dz}. Under the assumption that DPS dominates, the estimated effective cross-section corresponding to the measured one would have to be 2-3 times larger than for the J/psi case.
The cross-section measurements supports the hypothesis that the DPS mechanism plays a large role in the \mbox{2 x ($c \bar{c}$)} production at LHC energy, however the study of  $p_T$ spectra of charm hadrons from \mbox{2 x ($c \bar{c}$)}-processes questions the naive DPS paradigm. The  $p_T$ spectra expected to be similar between DPS \mbox{2 x ($c \bar c$)} processes and DPS ($c\bar{c}$) production, appear to be very different for $J/\psi$ + open charm hadron production compared with the single meson production~\cite{Aaij:1333554}. 
The whole pattern of  $p_T$ spectra is difficult to interpret when one single dominant mechanism is assumed. To explain all the investigated distributions, other mechanisms, e.g. charm sea-quarks, need also to be taken into account and could significantly contribute to the measured cross-section. However the sea-quarks approach is affected by large uncertainties and does not provide predictions for the kinematical distributions.  
Results on the $Z+$jets cross-section measurements of LHCb~\cite{ZjetLHCb}, nicely consistent with the SPS picture, confirm the difficulty of interpreting the experimental data using the available models.

ALICE studies\cite{Abelev:2012rz} the production of $J/\psi$ mesons as a function of the charged particle multiplicity in proton-proton collisions at $\sqrt{s} =$ 7.0 TeV. The $J/\psi$ particles are reconstructed using the invariant mass distribution of electron-positron pairs measured in the central rapidity region ($|y| <$ 0.9) as well as of muon pairs of opposite charge measured in the forward region (2.5 $< y <$ 4). Both these measurements show independently that the $J/\psi$ yield grows approximately linearly as a function of the charged particle multiplicity. A possible interpretation of these results is that the increase of the $J/\psi$ yield as a function of the charged particle multiplicity is due to MPI. Indeed, with increasing charged particle multiplicity resulting in the selection of events with a high number of multiple parton interactions, the yield of the $J/\psi$ is also expected to increases.

The DPS plans of ATLAS and CMS comprise a rich set of $\sigma_{eff}$ differential measurements in different final states and at different $\sqrt{s}$. Indeed, understanding the process dependency of $\sigma_{eff}$ is probably the most important task in the ``cahier de charges'' for the DPS research line at the LHC. However, for the time being both ATLAS and CMS focus on the benchmark inclusive measurements in final states with heavy bosons + jets, which aim at consolidating the DPS observation at the LHC and at comparing the inclusive $\sigma_{eff}$ measurement with the Tevatron figures.

ATLAS reports preliminary results \cite{Aad:2013bjm} on the production of W bosons in association with two jets in proton-proton collisions at a centre-of-mass energy of $\sqrt{s} = 7$ TeV. The DPS component is measured through the $p_T$ balance between the two jets and amounts to a fraction of 
0.08 $\pm$ 0.01 (stat.) $\pm$ 0.02 (sys.) for jets with $p_T >$ 20 GeV and rapidity 
$|y| <$ 2.8. This corresponds to a measurement of the effective area parameter for hard double-parton interactions of $\sigma_{eff} =$ 15 $\pm$ 3 (stat.) $^{+5}_{-3}$ (sys.) mb. The result turns out to be in agreement with the Tevatron measurements on 3~jet$ + \gamma$ topologies. 

The CMS contribution to the workshop  concentrates on simulation studies in the same channel which demonstrate the importance of the systematic effects arising from the considered DPS observable and from the model dependency for the simulation of the SPS background. To date only the soft MPI part is tuned in the state of the art Parton Shower Monte Carlo generators, however these tunes tend to ÒpredictÓ $\sigma_{eff}$ values between 20 and 30 mb, i.e. the predicted hard MPI cross sections turn out to be around 2-3 times lower than their measured values. Of course the underestimation of the extra-jet production through MPI may tend to be compensated by an excessive amount of radiation, i.e. an unphysical description of the inclusive multi-jet backgrounds, which may compromise other SM measurements, artificially increasing the systematic uncertainties connected to radiation. However one should also take into account the possibility that SPS processes generated by these Parton Shower Monte Carlo generators do not cover in a adequate way corners of the phase space, which are populated by DPS events, for example direct photon events having the 2-nd and the 3-rd jet with back-to-back configuration in the $R-\phi$ plane. In this case the fits would tend to overestimate the DPS rates and the measurements would report an artificially lower value of $\sigma_{eff}$. 

\subsubsection{Conclusions and future steps}

Overall it is prudent to wait for further DPS results from the LHC in order to say that we have robust DPS results.

Given the present conceptual and technical difficulties to model both the DPS signals and the SPS backgrounds, more  experimental results are needed,  even if the interpretation of the data will be left for a subsequent iteration, i.e. to publish the differential measurement of the cross sections in terms of DPS sensitive observables. 

At the same time, a striking evidence or rejection of the DPS formalism could be provided studying channels characterized by a  good separation between DPS signal and SPS background, focusing in particular on processes without jets, for example on the production of same sign W pairs \cite{Gaunt:2010pi}.

The production of two equal sign W bosons via SPS is a higher order process in the Standard Model, where two equal sign W bosons can be produced only in association with two jets. For such final state the DPS and SPS contributions turn out to be comparable in magnitude at the LHC, with the DPS process dominating at low  $p_T$ of the produced W. Assuming the current estimations for $\sigma_{eff}$, the cross section of DPS same sign W production with subsequent leptonic decay of the W is O(10fb). Considering the fact that the SPS background turns out to be reducible, this channel should be considered a benchmark to provide DPS evidence at the LHC.

The same signature can be adopted to study the DPS processes in pPb and PbPb interactions where, due to the large transverse parton densities, the number of MPI (and corresponding DPS) are sizably enhanced with respect to the pp case and which, due to the increased complexity of the target, allow access to even deeper properties of the non perturbative hadronic structure~\cite{dEnterria:2013ck}.

\makeatletter{}\subsection{Theoretical Discussion}
{\it Author:}  D. Treleani \\

\subsubsection{What are the Double Parton Scatterings?}
Hadrons are extended objects and, in collisions at large $p_t$, the interactions between hadronic constituents are localized in regions much smaller as compared to the hadronic dimension. If one considers hadronic collisions in a regime where hard interactions between hadronic constituents are not rare and rather represent a finite fraction of the total inelastic cross section, one may thus expect that there will be cases where the hard component of the hadronic interaction is not connected. When looking at DPS, one is looking at the simplest case of a disconnected hard interaction.

\subsubsection{Which are the general features of a DPS process? }
To have a DPS, the incoming flux of the partons, active for large momentum exchange processes, must be very large, which implies very large c.m. energies and low values of the incoming fractional momenta. Since the two hard interactions are localized in two different regions in transverse space, the DPS cross section has a much stronger dependence on the incoming parton flux, as compared to the more familiar case of a hard process originated in a connected region. Differently with usual large $p_t$ processes, a DPS is characterized by a non-perturbative scale, related to the typical transverse distance between the interaction regions. A consequence is that the DPS cross section decreases much faster as a function of transverse energy. The final state partons generated by a hard interaction are highly correlated in transverse momenta and in rapidity. Disconnected hard interactions are thus expected to generate various groups of highly correlated partons in the final state, while the correlations between partons belonging to different groups should be weak. 

The dependence on the incoming parton flux and on transverse energy and the correlations in the final state are therefore the main features, which characterize the disconnected hard interactions.

These general features of the DPS cross section are summarized in eq.(\ref{sdouble}).
When hard interactions are rare, the probability to have the process $B$ in a inelastic interaction is given by the ratio $\sigma_B/\sigma_{inel}$. Once the process $A$ takes place, the probability to have the process $B$ in the same inelastic interaction is of course different. It can anyway be always written as $\sigma_B/\sigma_{eff}$, where $\sigma_{eff}$ plays {\it effectively} the role played by the inelastic cross section in the unbiased case. The measured values of $\sigma_{eff}$ are much smaller (roughly by a factor 4) as compared with $\sigma_{inel}$, which is a clear indication of the presence of important correlations in the hadron structure. Notice that, although $\sigma_{eff}$ is related to the transverse distance between the two interaction $A$ and $B$, it does not give a direct measure of the transverse correlation between partons and therefore it cannot be understood as the {\it effective transverse interaction area}. Such identification would only be possible in absence of partonic correlations in fractional momenta.

Eq.(\ref{sdouble}) has shown to be able to describe the experimental results of the direct search of double parton collisions in rather different kinematical regimes~\cite{Akesson:1986iv, Abe:1997xk, Abazov:2009gc, ATLAS} with a value of $\sigma_{eff}$ compatible with a universal constant, while the study of CDF~\cite{Abe:1997xk}, of the dependence of $\sigma_{eff}$ on the fractional momenta of the incoming partons, is again compatible with a value of $\sigma_{eff}$ independent of $x$. One should point out that these experimental results represent a far from trivial test of eq.(\ref{sdouble}). As already mentioned, the expression of $\sigma_{double}^{(A,B)}$ in eq.(\ref{sdouble}) has in fact a rather strong dependence on the incoming parton flux, very different from the dependence on the incoming parton flux of a single hard scattering process. \\

\subsubsection{Why to study DPS?}
A first reason of interest is that, in some cases, DPS can give a sizable background in channels of interest for the search of new physics~\cite{Maina:2009sj}. A second reason of interest is that DPS, and in general MPI, probing the hadron in different points contemporarily, allows a much deeper insight into the hadron structure. The non-perturbative input to DPS is in fact given by the Double Parton Distribution Functions, which have a straight connection with the correlations in the hadron structure. Through DPS one may therefore have access to unprecedented information on multi-parton correlations, which cannot be obtained by studying only Single Partonic Scatterings with large $p_t$ exchange.

Correlations have in fact generated a lot of attention. An interesting study on helicity dependent Double Parton Distributions and spin correlations in double Dell-Yan type processes ($W^{\pm}$, $Z$, $\gamma^*$) has been reported at the Workshop by Tomas Kasemets. Correlations of non perturbative origin are naturally expected to play a major role in DPS. On the other hand, the possibility of observing correlations induced by perturbative QCD has been a topic of much activity. As discussed by Jonathan Gaunt in his overview talk, color induced correlations for a pair of partons at finite transverse distances are Sudakov suppressed. Correlations induced by perturbative splitting might, on the other hand, play a more relevant role. When the two active partons in a hadron originate from a perturbative splitting, the resulting DPS is necessarily localized in a narrow region in transverse space, much smaller as compared to the hadronic dimension. Correlations induced by perturbative splitting would thus represent a characteristic signature of the interplay between a connected and a disconnected hard interaction.

The topic of Correlations induced by perturbative splitting is still a matter of debate. According to Yuri Dokshitzer and collaborators~\cite{Blok:2011bu}, the back to back jet kinematics, characteristic of DPS are either generated by $(2\to2)^2$ partonic processes, which lead to the expression of the cross section in Eq.(\ref{sdouble}), or by $3\to4$ parton processes, where two of the initial state partons, originated from the perturbative splitting of a single parton from the hadron wave function, interact with two partons from the second hadron. The double hard collision of two parton pairs, each of which originates from perturbative splitting, lacks the back-to-back enhancement and should not be considered as a contribution to the DPS. The expression of the DPS cross section should therefore be modified, introducing two additional terms corresponding to the two possible $3\to4$ parton processes~\cite{Blok:2011bu}. Both Dokshitzer and Strikman argue that the enhanced rate of double parton collisions measured experimentally (namely the small value of $\sigma_{eff}$) could be explained by the presence of these additional contributions~\cite{Blok:2012mw}. 

On the other hand, according with Alexandre Snigirev, in the double logarithm approximation, all components of the generalized double distribution functions have the factorization structure and contribute to the cross section for a DPS process with the same leading exponential terms, but with different weights (non-exponential factors). Double splitting diagrams contribute therefore to a single parton pair collision (multiplied by a $2\to 4$ hard subprocess), with the contribution of the factorization type where the formation of two parton branches (one to two splitting) takes place at low scales. On the other hand, when the splitting takes place everywhere during the evolution (with more or less equal probabilities), the double splitting contribution cannot be considered any more as a contribution to a single scattering interaction~\cite{Ryskin:2011kk, Ryskin:2012qx}.

Concerning the characteristic properties of the additional $3\to4$ parton processes, according to Dokshitzer and collaborators, the four large $p_t$ partons in the final state are characterized by an overall unbalance in transverse momenta, much smaller as compared to the typical unbalance of the $(2\to2)^2$ process. Another peculiar feature is the different dependence on the incoming fractional momenta, which is originated by the very different initial state flux in the $3\to4$ and $(2\to2)^2$ cases. The $3\to4$ contribution grows in fact rapidly with the incoming fractional momenta, while it dies out quickly at small $x$, where correlations of non-perturbative origin are expected to dominate. 

The presence and the quantitative relevance of the $3\to4$ processes, in the kinematical regime where DPS are observed, awaits experimental confirmation and study. The experimental identification of the specific correlations in transverse momenta, induced by the presence of the $3\to4$ processes, for different combinations of initial state fractional momenta, would allow to characterize quantitatively the importance of the $3\to4$ processes and thus the contribution of the interplay region between connected and disconnected hard interactions.   

Regarding the small $x$ region, Strikman points out that a new non-perturbative mechanism, due to diffractive processes, can play an important role and enhance the DPS cross section by a factor of about two\cite{Blok:2012mw}, while an indication, on the behavior of DPS at small $Q^2$, might come from the decrease of the $3 \to 4$ mechanism as a function of the hardness of the process. The trend is consistent with the D0 data and would suggest that, when going to exchanged momenta of the order of 3 - 4 GeV/c, $\sigma_{eff}$ might grow to values as big as 25 - 30 mb. On the other hand, when the small $x$ region is probed by DPS, in double \J production and in the production of two pairs of $c \bar c$ in $pp$ collisions at the LHC, the DPS contribution is consistently described by the conventional $(2\to2)^2$ mechanism, with the standard value $\sigma_{eff}=15$ mb~\cite{BaranovJJ,Baranov:2011ch,LMS2012,Maciula:2013kd}.

With respect to DPS at low momenta exchanged, in his talk on the Kinematic correlations in double \J production, Nikolay  Zotov points out that, while pairwise compensation of transverse momenta is one of the main characteristics of a DPS at large transverse momenta, when including initial state radiation effects 
(either in the form of $k_t$-dependent gluon distributions~\cite{BaranovJJ} or by simulating the parton showers in a phenomenological way ~\cite{Stirling1}) the original azimuthal correlations are washed out, with the result of making the SPS and DPS samples very similar, unless going at sufficiently high $p_T$ ($p_T > 6$ GeV/c at LHC energy), where
the production rates are very small. The single and double parton scattering contributions are more efficiently separated by looking at large rapidity difference events~\cite{Baranov:2012re}. In the case of the DPS mode the
distribution over $\Delta y$ is rather flat, while 
the leading order SPS contribution is localized inside the interval $|\Delta{y}|\le 2$ (and continues to fall down steeply with increasing $|\Delta{y}|$). The higher order contributions extending beyond these limits are heavily suppressed by the color algebra and do not constitute significant background for the DPS production.

Similar results are reported by Antoni Szczurek, in his talk on the production of two pairs of $c \bar c$ in proton-proton collisions at the LHC, where higher order corrections are taken into account by using unintegrated gluon distributions to evaluate the cross sections. The production rates
for different combinations of charmed mesons, as well as the differential distributions in $D^0 D^0$ invariant
mass and the azimuthal correlations between two $D^0$ mesons, have been calculated\cite{LMS2012,SS2012,Maciula:2013kd} and, 
within large theoretical uncertainties, using however the constant value $\sigma_{eff}=15$ mb, the predicted DPS cross sections 
are found to be fairly similar to the cross sections measured recently 
by the LHCb collaboration~\cite{Aaij:2012dz}.

The issue of a more efficient strategy to select DPS events is discussed by Ed Berger in his talk on DPS contributions in $Wbb$ and $Zbb$ production at the LHC. Ed Berger points out that in most single-variable kinematic distributions, the DPS contribution is not distinctive enough to be picked out of the large SPS and background contributions.   However, there are signature kinematic variables that exploit the $2\to2$ parton nature of the underlying 
DPS subprocesses, such as approximate balance in transverse momentum and azimuthal angle correlations.  {\em Distributions of events in two-dimensional plots of such variables show distinct regions in which the DPS contribution dominates.}  In these well defined regions of phase space, the signal significance of the DPS contribution relative to the SPS and background contributions can be quite large~\cite{Berger:2011ep, Berger:2009cm}.

Obviously Double (and more in general Multiple) Parton Scatterings are more abundant in reactions with nuclei. On the other hand in many cases, in high energy nucleus-nucleus collisions, a disconnected hard interaction may simply correspond to hard interactions taking place between different pairs of nucleons. In his talk Urs Widemann points out that it may be particularly instructive to study DPS in proton-nucleus collisions. In $pA$ collisions, in a regime where non additive corrections to the nuclear parton distributions are small, DPS originate either from collisions with a single active target nucleon or from collisions with two different active target nucleons. While the first contribution does not add much to the information already available form a DPS on a isolated nucleon, the second contribution has the peculiar property of {\it filtering out} longitudinal correlations in the proton. Given the much larger radius of the nucleus, in the latter case the relative transverse distance between the interacting pairs does not play in fact any relevant role~\cite{Blok:2012jr}. By selecting the contribution to DPS with two active target nucleons, one will hence have direct access to the longitudinal correlations of the hadron structure, while the contributions of the transition region, from a connected to a disconnected hard interaction, are expected to be sizably reduced as compared to the $pp$ case.    \\

In summary, the main points emerged, in the theory contributions to the DPS Session, are the following:
\begin{itemize}
\item The $(2\to2)^2$ mechanism, with essentially the same value of $\sigma_{eff}$, reproduces the observed rates of DPS in a wide variety of kinematical regimes. Ranging, at the LHC, from the production of two pairs of $c \bar c$ to the production of $W +2j$.
\item DPS is a process where the incoming flux of partons is the square of the incoming flux in a Single Parton Scattering process, while the initial values of fractional momenta of each partonic collisions are experimentally accessible quantities. The experimental study of the dependence on the initial fractional momenta would thus be a valuable tool to help identifying DPS processes.
\item There is not a sharp separation between connected and disconnected components of the hard interaction and the transition region might give an appreciable contribution to the observed cross sections. From the theory side, the issue is still a matter of debate and might require additional thoughts. The experimental search of the events, with the peculiar signatures of the $3\to4$ processes, would allow to quantify the size of the contribution of the transition region and to identify the correspondingly relevant kinematical domain.
\item  By representing the DPS contribution with the conventional $(2\to2)^2$ mechanism, double \J production  and the production of two pairs of $c \bar c$, in proton-proton collisions at the LHC, are successfully described, including the production rates for different combinations of charmed mesons. In such a case, DPS events are dominantly generated with a rather small $p_t$, where correlations in transverse momenta are lost and the use of $k_t$ dependent parton distributions becomes an essential tool. An effective way to separate DPS events is to look at correlations in rapidity. 
\item On the other hand DPS events are characterized by specific correlations in different variables, so an efficient strategy to identify the contributions due to DPS will be to look at the distribution of events in two or more of those variables. 
\item An important issue concerns DPS in $pA$ collisions, which, providing an additional independent handle to learn about multi-parton correlations, can provide a lot of interesting information. In this case the hadron structure enters in fact in a peculiar way, as the contribution to DPS with two active nucleons depends only on the longitudinal correlations in the projectile proton.
\item One should finally remind the interesting experimental results on Triple Parton Scatterings (TPS): Both CDF and D0 report evidence of a sizable contamination of TPS in their samples of events with DPS, which represents a rather encouraging indication on the feasibility of the experimental study of TPS at the kinematical regimes presently available.
\end{itemize}

\section{MPI and MC development} 

{\it Author: } \\ J. Katzy, T. Rogers,  A. Si\'odmok, M. Strikman \\

\subsection{ Monte Carlos:}  A general overview of multiparton collisions in 
\MC s was provided by A. Si\'odmok.
The issue of the correct description of transverse matter distribution and 
optimal choice of a $p_T$ cutoff was discussed.  A comparison of different 
model treatments of color reconnections was also provided.  
\begin{itemize}
\item{\bf \sc Epos}\\
A new ansatz to describe MPI at LHC and in any hadronic collisions has been incorporated in the {\sc Epos} generator which was  presented including the {\sc Epos} LHC tune by T.Pierog. Particular emphasize  were given to the modeling of collective effects in {\sc Epos} which is inspired by commonly  used models to describe heavy ion collisions.  This model is implemented for the hadronisation  phase  as modifications to the string fragmentation depending on the string density in the final state. This model is able to describe all MB observables measured at LHC, 
in particular the charged particle pt spectra and the strange particle production  
which are not described very precisely described by any of the other models. 

In particular it was noted with interest that this model  gets the relation between charged particle multiplicity and mean transverse momentum right without implementing color reconnection as other models like {\sc Pythia} and {\sc Herwig} do.
\item{\bf \sc Herwig++}\\
Models with hard~\cite{Bahr:2008dy} (similar to  JIMMY~\cite{Butterworth:1996zw} package) 
and soft component~\cite{Bahr:2009ek} of multiple partonic interactions are available 
in  Herwig++\cite{Bahr:2008pv}. There were also details  studies on the color structure 
of multiple interactions leading to construction of two different color reconnection 
models~\cite{Gieseke:2012ft}. The current energy-dependent UE tunes (included in the last 
version of the generator~\cite{Arnold:2012fq}) are able to describe the existing primary charged 
particles UE data collected at different collider energies. 
\item{\bf \sc Sherpa}\\
A new model for minimum bias (MB) and  underlying event (UE) in {\sc Sherpa}~\cite{Gleisberg:2008ta}, SHRiMPS~\cite{Khoze:2010by}, has been presented by K. Zapp. The older, current default is similar to the Sj\"ostrand and Von Zijl model which is also used in Pythia. The new model is based on Khoze-Martin-Ryskin model, incorporating an eikonal ansatz to describe the proton as a superposition of diffractive eigenstates and should become the default in the future.
A tune and comparison to LHC data of the new model showed that agreement within 10\% with the MB rapidity distribution and the UE at 7 TeV can be reached. Larger deviations are observed at lower center-of-mass energy and in the MB differential distributions like the charged particle multiplicity and transverse momentum spectrum.

\item{\bf \sc Pythia8} Tunes \\
Tunes of the {\sc Pythia8} generator~\cite{Sjostrand:2007gs} done by the ATLAS Collaboration were presented by C. Wahrmund. The tunes used the model of x-dependent matter distribution and tuned to a various number of PDFs.  It was found that MB distributions are better described by LO PDFs. Reasonable description of the UE observables (which are more inclusive) could be reached with NLO PDFs. In total, a good description of UE and MB can be reached with {\sc Pythia8}, the agreement with the data is similar to recent  {\sc Pythia6} tunes, sometimes even slightly better.

\newpage
\item{\bf\sc Pythia8-mbr} \\
Robert Ciesielski and Konstantin (Dino) Goulianos
presented an option in {\sc pythia8} designed to simulate {\sc sd}, {\sc dd}, and {\sc dpe} processes based on the renormalization model for diffraction by Dino~\cite{RENORM} and implemented in {\sc Pythia8} by Robert since version 8.165~\cite{MBR_note}. At {\sc cdf}, all processes are well modeled by a stand-alone {\sc mbr} simulation based on a unitarized Regge-theory approach, employing inclusive nucleon {\sc pdf}'s and {\sc qcd} color factors. {\sc Renorm}~\cite{RENORM}  was updated for {\sc eds-2009} to include a novel prediction of the total $pp$ cross section~\cite{EDS2009_total}.  In {\sc Pythia8-mbr}, final state hadronization is performed by {\sc Pythia8}, tuned to reproduce final states in agreement with those of the original {\sc mbr} in which hadronization was based on a model producing only pions~\cite{Goulianos:1987ec}. 
\end{itemize}

\subsection{Measurements of Multiple Collisions:} 
\begin{itemize}
\item { \bf Strangeness production in UE and comparison to MC generators} \\
 T. Hreus presented for the CMS Collaboration a measurement of the strangeness production in the underlying event, 
extending the measurement of the same observables previously measured in MB. The observed characteristics of the UE activity is very similar to the one of primary charged particles. 
However, as in the 
case of MB, it was found that the strangeness production, both the multiplicity and the pt spectra, are  underestimated in all presented {\sc Pythia6} and {\sc Pythia8} tunes.

\item{\bf Common plots of the LPCC MB \& UE working group}\\
R. Field presented comparison of several MB\&UE "common plots" of the different LHC experiments. Observables in a common phase space were defined for these plots  by the LPCC MB\&UE working group in order to 
allow to compare the different experimental measurements. While the MB distributions agree 
very well between all experiments, the ALICE UE data are systematically a bit lower (but still within systematic uncertainty) than the ATLAS measurements, with CMS data lying in between.
Rick also presented a first analysis of the UE data from the Tevatron energy scan at 300 GeV, 900 GeV and 1.96 TeV. A slightly different center-of-mass energy dependence is found of the mean charged particle density and the mean charged particle $p_t$ sum density. A decent description of the data 
from 300 GeV up to 7 TeV with the {\sc Pythia6} tune Z1 was reached.
Comparing Tevatron and LHC data at the same center-of-mass energy is possible with the 900 GeV data set. 
A slightly higher (~4\%) plateau in the UE distribution of Tevatron was found in the leading $p_t$. A potential cause 
of this difference could the difference in the quark luminosities between proton-antiproton and proton-proton collision leading to  different sub-processes contributing to  charge particle production. 
The analysis of the full data set  is ongoing.
\item {\bf DPS} \\
Measurements of 
double hard collisions with $W$ plus two jet events at ATLAS were explained by I. Sadeh with a value of 
\begin{equation}
\sigma_{\rm eff} = 15 \pm 3\, \rm{(stat.)} \; {}^{+5}_{-3} \, \rm{(sys.)} \; {\rm mb} \nonumber 
\end{equation}
being reported.
A new analysis of CDF data was reported by M. Myska with a result of 
\begin{equation}
\sigma_{\rm eff} = 12.0 \pm 1.3^{+1.7}_{-2.3} \; {\rm mb} \, . \nonumber
\end{equation}
\end{itemize}

\subsection{ QCD and Modeling of Multiple Interactions}  
General overviews
of the theoretical description of multiple parton collisions within a parton 
model and perturbative QCD was provided by D. Treleani, J. Gaunt and Y. Dokshitzer.   
A description of multiple hard collisions in impact parameter space, and the relationship between the impact parameter description and partonic correlations, was provided by M. Strikman.  He emphasized that the models which assume that parton distributions are not correlated in the transverse coordinate and momentum and constrained by the data from hard exclusive processes leads to much larger $\sigma_{eff} \ge  32$ mb, than suggested by the data. 
Recent work on the angular modulations that can result from 
multiple hard collisions was explained by T. Kasemets.  The potential to extract measurements 
of double hard collisions from $W$ $bb$ events and $Z$ $bb$ events at the LHC was 
explained by E. Berger.

\subsection{To-Do List}
The list of steps needed for the future is 
characterized by the need to relate the methodology of 
\MC s to recent theoretical developments in the treatment 
of multiple collisions.  This includes the following:
\begin{itemize}
\item The value of $\sigma_{\textit{eff}}$ should serves as a constraint on the Monte Carlo models since the recent tunes of MPI models to the LHC data predict its value to be between $25-42$~mb~\cite{Gieseke:2011xy}. 
\item Incorporating perturbatively determined phenomena (such as the gluon generalized parton distribution for the transverse width of partonic matter in the proton as measured in exclusive hard processes and the angular correlations discussed by T. Kasements) into \MC\  descriptions.
\item Relating color reconnections to an underlying QCD description.
\end{itemize}

\section{Appendix} 
\makeatletter{}
 
\subsection{Extracting $\sigma_{eff}$ from the CDF $\gamma$ + 3 jets measurement \label{bahr}}

{\it Authors:} \\
M. B\"ahr, M. Myska, M. H. Seymour, A. Si\'odmok \\

The CDF measurement of double parton scattering~\cite{Abe:1997xk} used a non-standard 
definition of $\sigma_{eff}$ that makes this important quantity process dependent. Therefore,
the value provided by the experiment
\begin{equation}
 \sigma_{eff,CDF}  = (14.5\pm1.7^{+1.7}_{-2.3})~\mbox{mb}
\end{equation}
is not suitable for comparisons with other measurements or as input for theoretical calculations
or Monte Carlo models. The non-standard definition used by CDF has been pointed out and corrected for the first time by
Treleani in his publication~\cite{Treleani:2007gi}. Based on the theoretically-pure (parton-level) situation, \cite{Treleani:2007gi}
estimated the inclusive (process independent)
$
  \sigma_{eff}=10.3~\mbox{mb}.
$
This result would be correct under the assumption that CDF were able to uniquely identify 
and count the number of scatters in an event, which is certainly not the case.
In \cite{Bahr:2013gkj} we have considered CDF's event definition in more detail to provide an improved correction leading to
\begin{equation}
 \sigma_{eff} = (12.0 \pm 1.4 ^{+1.3}_{-1.5})~\mbox{mb}.
\end{equation} 
It is worth noting that both statistical and systematic uncertainties have decreased, since the additional uncertainty 
of our correction factor is much smaller than the avoided uncertainty stemming from the triple scattering removal done originally by CDF.

The obtained value of $\sigma_{eff}$ serves as a constraint on the Monte Carlo models since the recent tunes of MPI
models to the LHC data predict its value to be between $25-42$~mb~\cite{Gieseke:2011xy}. 
This inconsistency between theory and experiment indicates that the overlap function used in the MC models is oversimplified and should be improved, for example, 
by including $x$-dependence~\cite{Strikman:2011cx,Frankfurt:2010ea,Corke:2011yy}. This value can also help to understand 
the recent results from the LHCb experiment~\cite{Aaij:2012dz}(see page 23, Fig 10).  
The experimental results for $\sigma_{eff}$ extracted from the production of $J/\psi$ mesons together with an associated open
charm hadron and from double open charm hadron production are different by a factor of between two and three.

\subsection{EPOS3}
{\it Author: } \\
K. Werner, B. Guiot, Iu. Karpenko, T. Pierog \\

Inclusive cross sections are particularly simple, quantum interference
helps to provide simple formulas referred to as {}``factorization''.
Although factorization is widely used, strict mathematical proofs exist
only in very special cases, and certainly not for hadron production
in $pp$ scattering. To go beyond factorization and to formulate a
consistent multiple scattering theory is difficult. A possible solution
is Gribov's Pomeron calculus, which can be adapted to our language
by identifying Pomeron and parton ladder. Multiple scattering means
that one has contributions with several parton ladders in parallel. 

\begin{figure}\hskip 2cm 
\includegraphics[scale=0.25]{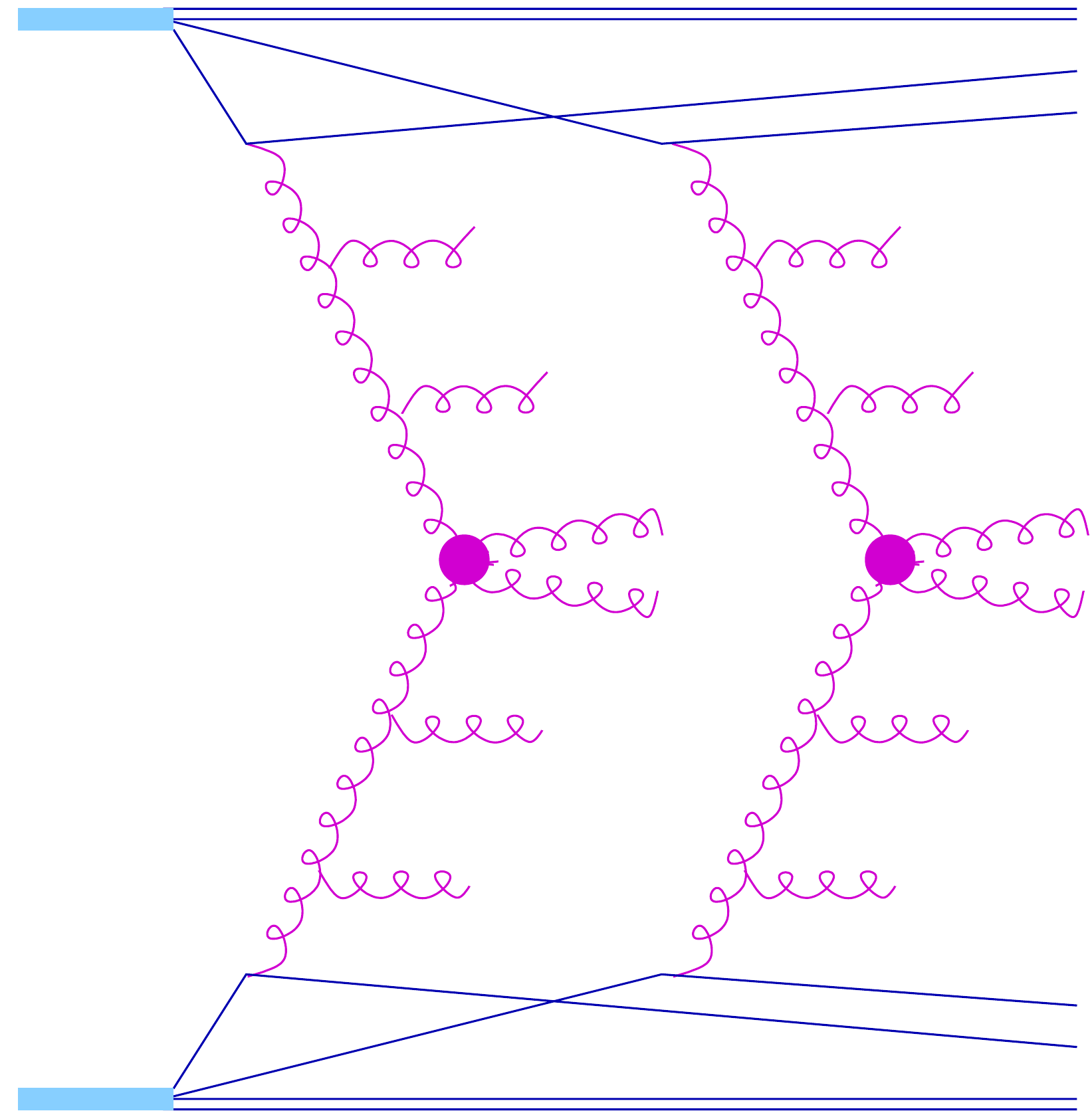}
 \vskip -5cm \hskip 7.5cm \includegraphics[angle=0,scale=0.3]{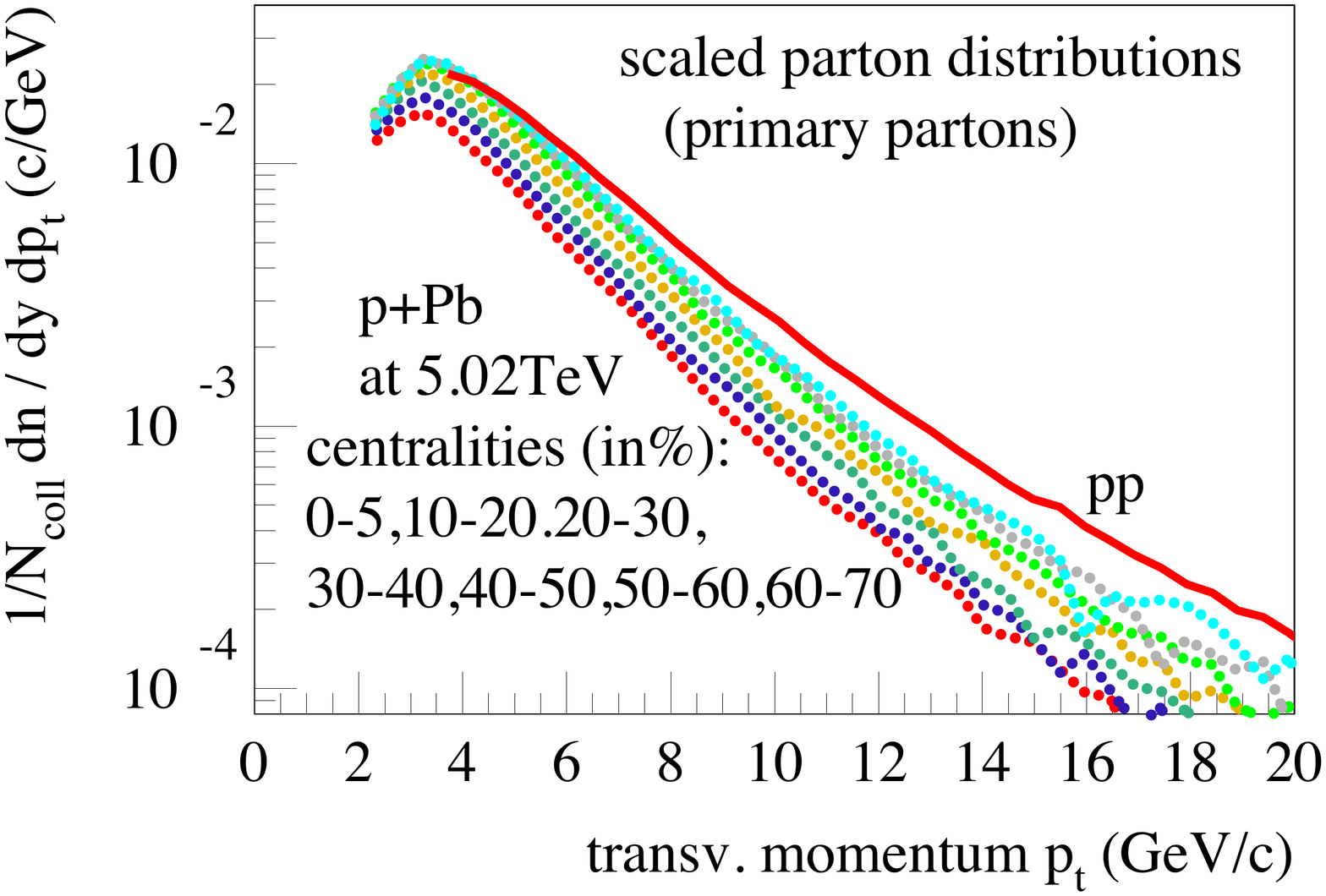}  \begin{picture}(0,0)
      \put(-350,+10.){{$(a)$}}
      \put(-100,+10){{$(b)$}}
    \end{picture}
\caption{$(a.)$ Schematic diagram for multiple scattering. $(b.)$ Scaled pPb cross section
with centrality increasing from top to bottom (bottom = most central)}
\label{Epos0}
\end{figure}

We indicated several years ago inconsistencies in this approach, proposing
an {}``energy conserving multiple scattering treatment''. The main
idea is simple: in case of multiple scattering, when it comes to calculating
partial cross sections for double, triple ... scattering, one has
to explicitly care about that the partons ladders which happen
to be parallel to each other (see Fig.\ref{Epos0}a)
\\
share the collision energy. This energy sharing has been implemented
in {\sc Epos}, which is a multiple scattering approach corresponding to
a marriage of Gribov-Regge theory and perturbative QCD (for details
see \cite{Drescher:2000ha,Werner:2010aa}). An
elementary scattering corresponds to a parton ladder, containing a
hard scattering calculable based on pQCD, including initial and final
state radiation. 

The energy sharing scheme generalizes straightforwardly to proton-nucleus
collisions. Here one expects (and observes) the so-called binary scaling,
which means that the scaled AA cross section (cross section divided
by the number of binary collisions) is equal to the proton-proton
one, \[
\frac{1}{N_{\mathrm{coll}}}\left.\frac{dn}{dp_{t}}\right|_{AA}=\left.\frac{dn}{dp_{t}}\right|_{pp}.\]
In Fig.\ref{Epos0}b, the simulation shows a strong violation of
this scaling, due to the imposed energy conservation.

\begin{figure}
\begin{center}
\includegraphics[angle=0,scale=0.35]{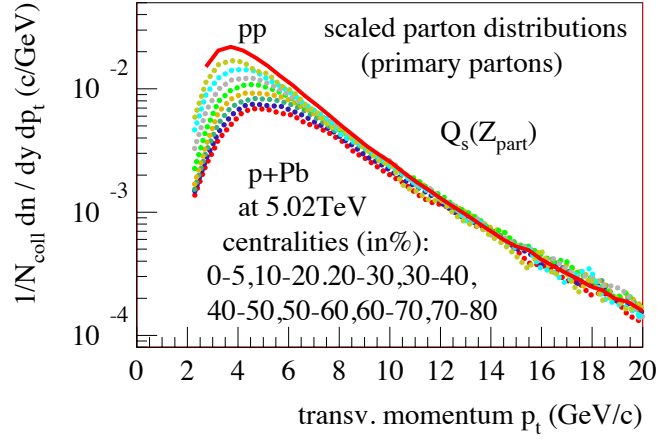}\end{center} 
\caption{New scaling procedure in {\sc Epos3}}
\label{Epos3}
\end{figure}

After many attempts, we finally found a solution to the problem: The
usual constant {}``soft'' scale $Q_{0}^{2}$ (the lower cutoff of
virtualities in the parton ladder) has to be replaced by a centrality
dependent saturation scale,

\noindent \[
Q_{s}^{2}=Q_{0}^{2}\,\bigg(1+B_{\mathrm{satur}}N_{\mathrm{part}}(i,j)\bigg),\]

\noindent where $N_{\mathrm{part}}(i,j)$ is the number of participating
nucleons connected to a given parton ladder between projectile nulceon
$i$ and target nucleon $j$. So each parton ladder has {}``its own''
saturation scale! This new procedure (implemented in {\sc Epos3}, with a
single free parameter $B_{\mathrm{satur}}$) allows to completely recover
binary scaling, see  Figure~\ref{Epos3}. 
So for the first time energy sharing is compatible with binary scaling
in a multiple scattering scheme.

\section*{Acknowledgements}
{The authors are grateful to everybody that contributed to the success of the session: speakers, fellow conveners and CERN workshop organizers. We are particularly  indebted to the speakers that contributed directly in writing this summary.


\begin{thebibliography}{10}
\newcommand{\enquote}[1]{``#1''}
\providecommand{\url}[1]{\texttt{#1}}
\providecommand{\urlprefix}{URL }
\providecommand{\eprint}[2][]{\url{#2}}

\bibitem{mpilhc2012}
\enquote{MPI@LHC 2012,} \\ \small
  \mbox{https://indico.cern.ch/conferenceOtherViews.py?view=standard\&confId=184925}.

\bibitem{Antchev:2013gaa}
G.~Antchev \emph{et~al.} (TOTEM Collaboration), \enquote{{Measurement of
  proton-proton elastic scattering and total cross-section at S**(1/2) =
  7-TeV},} \emph{Europhys.Lett.}, \textbf{101}(2013), 21002.

\bibitem{Antchev:2013iaa}
G.~Antchev \emph{et~al.} (TOTEM Collaboration),
  \enquote{{Luminosity-independent measurements of total, elastic and inelastic
  cross-sections at S**(1/2) = 7-TeV},} \emph{Europhys.Lett.},
  \textbf{101}(2013), 21004.

\bibitem{Antchev:1495764}
G~Antchev, \emph{et~al.}, \enquote{A luminosity-independent measurement of the
  proton-proton total cross-section at $\sqrt{s}$ = 8 TeV,} Technical Report
  TOTEM-2012-005. CERN-PH-EP-2012-354, CERN, Geneva, 2012.

\bibitem{Cudell:2002xe}
J.R. Cudell \emph{et~al.} (COMPETE Collaboration), \enquote{{Benchmarks for the
  forward observables at RHIC, the Tevatron Run II and the LHC},}
  \emph{Phys.Rev.Lett.}, \textbf{89}(2002), 201801, \eprint{hep-ph/0206172}.

\bibitem{Manohar:2012pe}
Aneesh~V. Manohar and Wouter~J. Waalewijn, \enquote{{What is Double Parton
  Scattering?}} \emph{Phys.Lett.}, \textbf{B713}(2012), 196.

\bibitem{Akesson:1986iv}
T.~Akesson \emph{et~al.} (Axial Field Spectrometer collaboration),
  \enquote{{Double parton scattering in $pp$ collisions at $\sqrt s =
  63$~GeV},} \emph{Z. Phys.}, \textbf{C34}(1987), 163.

\bibitem{Abe:1993rv}
F.~Abe \emph{et~al.} (CDF collaboration), \enquote{{Study of four jet events
  and evidence for double parton interactions in $p\bar{p}$ collisions at
  $\sqrt{s} = 1.8$ TeV},} \emph{Phys. Rev.}, \textbf{D47}(1993), 4857.

\bibitem{Sjostrand:1986ep}
Torbjorn Sjostrand and Maria van Zijl, \enquote{Multiple parton-parton
  interactions in an impact parameter picture,} \emph{Phys. Lett.},
  \textbf{B188}(1987), 149.

\bibitem{Abe:1997xk}
F.~Abe \emph{et~al.} (CDF collaboration), \enquote{{Double parton scattering in
  $\bar{p}p$ collisions at $\sqrt{s} = 1.8 $TeV},} \emph{Phys.Rev.},
  \textbf{D56}(1997), 3811.

\bibitem{Abazov:2009gc}
V.M. Abazov \emph{et~al.} (D0 collaboration), \enquote{{Double parton
  interactions in $\gamma + $~3~jet events in $p \bar p$ collisions
  $\sqrt{s}=1.96$ TeV},} \emph{Phys.Rev.}, \textbf{D81}(2010), 052012,
  \eprint{0912.5104}.

\bibitem{Treleani:2007gi}
D.~Treleani, \enquote{{Double parton scattering, diffraction and effective
  cross section},} \emph{Phys.Rev.}, \textbf{D76}(2007), 076006,
  \eprint{0708.2603}.

\bibitem{Aaij:2011yc}
R.~Aaij \emph{et~al.} (LHCb collaboration), \enquote{{Observation of $J/\psi$
  pair production in $pp$ collisions at $\sqrt{s}=7 TeV$},} \emph{Phys.Lett.},
  \textbf{B707}(2012), 52, \eprint{1109.0963}.

\bibitem{Aaij:2012dz}
R.~Aaij \emph{et~al.} (LHCb collaboration), \enquote{{Observation of double
  charm production involving open charm in pp collisions at $\sqrt{s}$=7 TeV},}
  \emph{JHEP}, \textbf{1206}(2012), 141, \eprint{1205.0975}.

\bibitem{Kiselev:1988mc}
V.V. Kiselev, A.K. Likhoded, S.R. Slabospitsky, and A.V. Tkabladze,
  \enquote{{Hadronic production of $\psi$ particle with large $M(\psi \psi$)},}
  \emph{Yad.Fiz.}, \textbf{49}(1989), 1681.

\bibitem{Qiao:2009kg}
Cong-Feng Qiao, Li-Ping Sun, and Peng Sun, \enquote{{Testing charmonium
  production mechanism via polarized $J/\psi$ pair production at the LHC},}
  \emph{J.Phys.}, \textbf{G37}(2010), 075019, \eprint{0903.0954}.

\bibitem{Novoselov:2011ff}
Alexey Novoselov, \enquote{{Double parton scattering as a source of quarkonia
  pairs in LHCb},} (2011), \eprint{1106.2184}.

\bibitem{Berezhnoy:1998aa}
A.V. Berezhnoy, V.V. Kiselev, A.K. Likhoded, and A.I. Onishchenko,
  \enquote{{Doubly charmed baryon production in hadronic experiments},}
  \emph{Phys.Rev.}, \textbf{D57}(1998), 4385, \eprint{hep-ph/9710339}.

\bibitem{Aaij:1333554}
R.~Aaij \emph{et~al.} (LHCb collaboration), \enquote{Measurement of $J/ \psi$
  production in $pp$ collisions at $\sqrt{s}$ = 7 TeV,} \emph{Eur. Phys. J. C},
  \textbf{71}(2011), 1645.

\bibitem{ZjetLHCb}
LHCb collaboration, \enquote{{Measurement of jet production in Z$/\gamma^* \to
  \mu^+ \mu^-$ events at LHCb in $\sqrt{s}=7$ TeV pp collisions},}
  \emph{LHCb-CONF-2012-016}.

\bibitem{Abelev:2012rz}
B.~Abelev \emph{et~al.} (ALICE Collaboration), \enquote{{$J/\psi$ Production as
  a Function of Charged Particle Multiplicity in $pp$ Collisions at $\sqrt{s} =
  7$ TeV},} \emph{Phys.Lett.}, \textbf{B712}(2012), 165, \eprint{1202.2816}.

\bibitem{Aad:2013bjm}
Georges Aad \emph{et~al.} (ATLAS Collaboration), \enquote{{Measurement of hard
  double-parton interactions in $ W+ 2$ jet events at $\sqrt(s)=7$ TeV with the
  ATLAS detector},} \emph{New J.Phys.}, \textbf{15}(2013), 033038,
  \eprint{1301.6872}.

\bibitem{Gaunt:2010pi}
Jonathan~R. Gaunt, Chun-Hay Kom, Anna Kulesza, and W.~James Stirling,
  \enquote{{Same-sign W pair production as a probe of double parton scattering
  at the LHC},} \emph{Eur.Phys.J.}, \textbf{C69}(2010), 53, \eprint{1003.3953}.

\bibitem{dEnterria:2013ck}
David d'Enterria and Alexander~M. Snigirev, \enquote{{Enhanced J/Psi production
  from double parton scatterings in nucleus-nucleus collisions at the Large
  Hadron Collider},} (2013), \eprint{1301.5845}.

\bibitem{ATLAS}
Georges Aad \emph{et~al.} (ATLAS Collaboration), \enquote{{Measurement of hard
  double-parton interactions in W+ 2 jet events at sqrt(s)=7 TeV with the ATLAS
  detector},} \emph{New J.Phys.}, \textbf{15}(2013), 033038,
  \eprint{1301.6872}.

\bibitem{Maina:2009sj}
Ezio Maina, \enquote{{Multiple Parton Interactions in Z+4j, W+- W+- + 0/2j and
  W+ W- + 2j production at the LHC},} \emph{JHEP}, \textbf{0909}(2009), 081,
  \eprint{0909.1586}.

\bibitem{Blok:2011bu}
B.~Blok, Yu. Dokshitser, L.~Frankfurt, and M.~Strikman, \enquote{{pQCD physics
  of multiparton interactions},} \emph{Eur.Phys.J.}, \textbf{C72}(2012), 1963,
  \eprint{1106.5533}.

\bibitem{Blok:2012mw}
B.~Blok, Yu. Dokshitzer, L.~Frankfurt, and M.~Strikman, \enquote{{Origins of
  Parton Correlations in Nucleon and Multi-Parton Collisions},} (2012),
  \eprint{1206.5594}.

\bibitem{Ryskin:2011kk}
M.G. Ryskin and A.M. Snigirev, \enquote{{A Fresh look at double parton
  scattering},} \emph{Phys.Rev.}, \textbf{D83}(2011), 114047,
  \eprint{1103.3495}.

\bibitem{Ryskin:2012qx}
M.G. Ryskin and A.M. Snigirev, \enquote{{Double parton scattering in double
  logarithm approximation of perturbative QCD},} \emph{Phys.Rev.},
  \textbf{D86}(2012), 014018, \eprint{1203.2330}.

\bibitem{BaranovJJ}
S.P. Baranov, \enquote{{Pair production of $J/\psi$ mesons in the
  $k_t$-factorization approach},} \emph{Phys.Rev.}, \textbf{D84}(2011), 054012.

\bibitem{Baranov:2011ch}
S.P. Baranov, A.M. Snigirev, and N.P. Zotov, \enquote{{Double heavy meson
  production through double parton scattering in hadronic collisions},}
  \emph{Phys.Lett.}, \textbf{B705}(2011), 116, \eprint{1105.6276}.

\bibitem{LMS2012}
Marta Luszczak, Rafal Maciula, and Antoni Szczurek, \enquote{{Production of two
  $c \bar c$ pairs in double-parton scattering},} \emph{Phys.Rev.},
  \textbf{D85}(2012), 094034, \eprint{1111.3255}.

\bibitem{Maciula:2013kd}
Rafal Maciula and Antoni Szczurek, \enquote{{Production of $c \bar c c \bar c$
  in double-parton scattering within $k_{t}$-factorization approach --
  meson-meson correlations},} (2013), \eprint{1301.4469}.

\bibitem{Stirling1}
C.H. Kom, A.~Kulesza, and W.J. Stirling, \enquote{{Pair Production of J/psi as
  a Probe of Double Parton Scattering at LHCb},} \emph{Phys.Rev.Lett.},
  \textbf{107}(2011), 082002, \eprint{1105.4186}.

\bibitem{Baranov:2012re}
S.P. Baranov, A.M. Snigirev, N.P. Zotov, A.~Szczurek, and W.~Schafer,
  \enquote{{Interparticle correlations in the production of $J/\psi$ pairs in
  proton-proton collisions},} \emph{Phys.Rev.}, \textbf{D87}(2013), 034035,
  \eprint{1210.1806}.

\bibitem{SS2012}
Wolfgang Schafer and Antoni Szczurek, \enquote{{Production of two $c \bar c$
  pairs in gluon-gluon scattering in high energy proton-proton collisions},}
  \emph{Phys.Rev.}, \textbf{D85}(2012), 094029, \eprint{1203.4129}.

\bibitem{Berger:2011ep}
Edmond~L. Berger, C.B. Jackson, Seth Quackenbush, and Gabe Shaughnessy,
  \enquote{{Calculation of W b bbar Production via Double Parton Scattering at
  the LHC},} \emph{Phys.Rev.}, \textbf{D84}(2011), 074021, \eprint{1107.3150}.

\bibitem{Berger:2009cm}
Edmond~L. Berger, C.B. Jackson, and Gabe Shaughnessy, \enquote{{Characteristics
  and Estimates of Double Parton Scattering at the Large Hadron Collider},}
  \emph{Phys.Rev.}, \textbf{D81}(2010), 014014, \eprint{0911.5348}.

\bibitem{Blok:2012jr}
Boris Blok, Mark Strikman, and Urs~Achim Wiedemann, \enquote{{Hard four-jet
  production in pA collisions},} (2012), \eprint{1210.1477}.

\bibitem{Bahr:2008dy}
Manuel Bahr, Stefan Gieseke, and Michael~H. Seymour, \enquote{{Simulation of
  multiple partonic interactions in Herwig++},} \emph{JHEP},
  \textbf{0807}(2008), 076, \eprint{0803.3633}.

\bibitem{Butterworth:1996zw}
J.M. Butterworth, Jeffrey~R. Forshaw, and M.H. Seymour, \enquote{{Multiparton
  interactions in photoproduction at HERA},} \emph{Z.Phys.},
  \textbf{C72}(1996), 637, \eprint{hep-ph/9601371}.

\bibitem{Bahr:2009ek}
Manuel Bahr, Jonathan~M. Butterworth, Stefan Gieseke, and Michael~H. Seymour,
  \enquote{{Soft interactions in Herwig++},} (2009), 239, \eprint{0905.4671}.

\bibitem{Bahr:2008pv}
M.~Bahr, \emph{et~al.}, \enquote{{Herwig++ Physics and Manual},}
  \emph{Eur.Phys.J.}, \textbf{C58}(2008), 639, \eprint{0803.0883}.

\bibitem{Gieseke:2012ft}
Stefan Gieseke, Christian Rohr, and Andrzej Siodmok, \enquote{{Colour
  reconnections in Herwig++},} \emph{Eur.Phys.J.}, \textbf{C72}(2012), 2225,
  \eprint{1206.0041}.

\bibitem{Arnold:2012fq}
K.~Arnold, \emph{et~al.}, \enquote{{Herwig++ 2.6 Release Note},} (2012),
  \eprint{1205.4902}.

\bibitem{Gleisberg:2008ta}
T.~Gleisberg, \emph{et~al.}, \enquote{{Event generation with SHERPA 1.1},}
  \emph{JHEP}, \textbf{0902}(2009), 007, \eprint{0811.4622}.

\bibitem{Khoze:2010by}
V.A. Khoze, F.~Krauss, A.D. Martin, M.G. Ryskin, and K.C. Zapp,
  \enquote{{Diffraction and correlations at the LHC: Definitions and
  observables},} \emph{Eur.Phys.J.}, \textbf{C69}(2010), 85,
  \eprint{1005.4839}.

\bibitem{Sjostrand:2007gs}
Torbjorn Sjostrand, Stephen Mrenna, and Peter~Z. Skands, \enquote{{A Brief
  Introduction to PYTHIA 8.1},} \emph{Comput.Phys.Commun.}, \textbf{178}(2008),
  852, \eprint{0710.3820}.

\bibitem{RENORM}
K.~Goulianos, \enquote{{Hadronic diffraction: Where do we stand?}} (2004), 251,
  \eprint{hep-ph/0407035}.

\bibitem{MBR_note}
R.~Ciesielski and K.~Goulianos, \enquote{{MBR Monte Carlo Simulation in
  PYTHIA8},} (2012), \eprint{1205.1446}.

\bibitem{EDS2009_total}
K.~Goulianos, \enquote{Diffractive and total $pp$ cross sections at {\sc lhc},}
  in Mario Deile, ed., \enquote{{Elastic and Diffractive Scattering.
  Proceedings, 13th International Conference, Blois Workshop, CERN, Geneva,
  Switzerland, June 29-July 3, 2009},} 2010, \eprint{1002.3527}.

\bibitem{Goulianos:1987ec}
K.~Goulianos, \enquote{{A new statistical description of hadronic and $e^+e^-$
  multiplicity distributions},} \emph{Phys.Lett.}, \textbf{B193}(1987), 151.

\bibitem{Gieseke:2011xy}
S.~Gieseke, C.A. R{\"o}hr, and A.~Si\'odmok, \enquote{{Multiple Partonic
  Interaction Developments in Herwig++},} (2011), \eprint{1110.2675}.

\bibitem{Bahr:2013gkj}
Manuel {B\"ahr}, Miroslav Myska, Michael~H. Seymour, and Andrzej Siodmok,
  \enquote{{Extracting $\sigma_{eff}$ from the CDF $\gamma+3$ jets
  measurement},} \emph{JHEP}, \textbf{1303}(2013), 129, \eprint{1302.4325}.

\bibitem{Strikman:2011cx}
Mark Strikman, \enquote{{Transverse Nucleon Structure and Multiparton
  Interactions},} \emph{Acta Phys.Polon.}, \textbf{B42}(2011), 2607,
  \eprint{1112.3834}.

\bibitem{Frankfurt:2010ea}
L.~Frankfurt, M.~Strikman, and C.~Weiss, \enquote{{Transverse nucleon structure
  and diagnostics of hard parton-parton processes at LHC},} \emph{Phys.Rev.},
  \textbf{D83}(2011), 054012, \eprint{1009.2559}.

\bibitem{Corke:2011yy}
Richard Corke and Torbjorn Sjostrand, \enquote{{Multiparton Interactions with
  an x-dependent Proton Size},} \emph{JHEP}, \textbf{1105}(2011), 009,
  \eprint{1101.5953}.

\bibitem{Drescher:2000ha}
H.J. Drescher, M.~Hladik, S.~Ostapchenko, T.~Pierog, and K.~Werner,
  \enquote{{Parton based Gribov-Regge theory},} \emph{Phys.Rept.},
  \textbf{350}(2001), 93, \eprint{hep-ph/0007198}.

\bibitem{Werner:2010aa}
K.~Werner, Iu. Karpenko, T.~Pierog, M.~Bleicher, and K.~Mikhailov,
  \enquote{{Event-by-Event Simulation of the Three-Dimensional Hydrodynamic
  Evolution from Flux Tube Initial Conditions in Ultrarelativistic Heavy Ion
  Collisions},} \emph{Phys.Rev.}, \textbf{C82}(2010), 044904,
  \eprint{1004.0805}.

\end{thebibliography}
\end{document}